# Observation of topological electronic structure in quasi-1D superconductor TaSe$_3$

Cheng Chen[1,2,3*], Aiji Liang[1,2*], Shuai Liu[1*], Simin Nie[4], Junwei Huang[5], Meixiao Wang[1,2‡], Yiwei Li[1,6], Ding Pei[6], Haifeng Yang[1,2], Huijun Zheng[1], Yong Zhang[7], Donghui Lu[8], Makoto Hashimoto[8], Alexei Barinov[9], Chris Jozwiak[3], Aaron Bostwick[3], Eli Rotenberg[3], Xufeng Kou[2,7], Lexian Yang[10,11], Yanfeng Guo[1], Zhijun Wang[12,13], Hongtao Yuan[5], Zhongkai Liu[1,2†], Yulin Chen[1,2,6,10§]

[1]*School of Physical Science and Technology, ShanghaiTech University, Shanghai 201210, P. R. China*
[2]*ShanghaiTech Laboratory for Topological Physics, Shanghai 201210, P. R. China*
[3]*Advanced Light Source, Lawrence Berkeley National Laboratory, Berkeley, CA 94720, USA*
[4]*Department of Materials Science and Engineering, Stanford University, Stanford, CA 94305, USA*
[5]*National Laboratory of Solid State Microstructures, College of Engineering and Applied Sciences, and Collaborative Innovation Center of Advanced Microstructures, Nanjing University, Nanjing 210093, China*
[6]*Department of Physics, University of Oxford, Oxford, OX1 3PU, United Kingdom*
[7]*School of Information Science and Technology, ShanghaiTech University, Shanghai 201210, P. R. China*
[8]*Stanford Synchrotron Radiation Lightsource, SLAC National Accelerator Laboratory, Menlo Park, California 94025, USA*
[9]*Elettra-Sincrotrone Trieste, Trieste, Basovizza 34149, Italy*
[10]*State Key Laboratory of Low Dimensional Quantum Physics, Department of Physics, Tsinghua University, Beijing 100084, China*
[11]*Frontier Science Center for Quantum Information, Beijing 100084, P. R. China*
[12]*Beijing National Laboratory for Condensed Matter Physics, and Institute of Physics, Chinese Academy of Sciences, Beijing 100190, China*
[13]*University of Chinese Academy of Sciences, Beijing 100049, China*

*These authors contributed equally to this work

‡wangmx@shanghaitech.edu.cn
†liuzhk@shanghaitech.edu.cn
§ yulin.chen@physics.ox.ac.uk (Lead Contact)

**Summary:**

**Topological superconductors (TSCs)，with the capability to host Majorana bound states that can lead to non-Abelian statistics and application in quantum computation，have been one of the most intensively studied topics in condensed matter physics recently. Up to date, only a few compounds have been proposed as candidates of intrinsic TSCs, such as doped topological insulator $Cu_xBi_2Se_3$ and iron-based superconductor $FeTe_{0.55}Se_{0.45}$. Here, by carrying out synchrotron and laser based angle-resolved photoemission spectroscopy (ARPES), we systematically investigated the electronic structure of a quasi-1D superconductor TaSe$_3$, and identified the nontrivial topological surface states. In**

**addition, our scanning tunneling microscopy (STM) study revealed a clean cleaved surface with a persistent superconducting gap, proving it suitable for further investigation of potential Majorana modes. These results prove TaSe$_3$ as a stoichiometric TSC candidate that is stable and exfoliable, therefore a great platform for the study of rich novel phenomena and application potentials.**

## I. INTRODUCTION

A topological superconductor (TSC) is a new type of superconductor with non-trivial topology in its bulk electronic structures that can lead to the emergence of Majorana bound states (MBSs) or Majorana Fermions (MFs) within its bulk superconducting gap[1,2]. The MBSs or MFs are collective excitations that are their own antiparticles with potential applications in topological quantum computation, which has inspired intensive investigations[3,4,5]. To realize the topological superconductivity, one can either realize a superconductor with intrinsic odd-parity pairing symmetry (e.g. as proposed in Sr$_2$RuO$_4$[6,7] and Cu$_x$Bi$_2$Se$_3$[8,9,10,11]), or construct heterostructures with interfaces between a superconductor and a topological insulator[12,13], or a semiconductor with strong spin-orbit interaction[14,15], or magnetic atom chains[16,17,18]. Due to the difficulty in identifying intrinsic odd-parity superconducting pairing, much recent effort has been made along the latter approach and yielded many exciting results, such as the observation of MBSs/MFs in 1D (e.g. InSb or InAs nanowires on superconductor[19,20]) and 2D (e.g. Bi$_2$Te$_3$ on superconductors[21,22] and Cr doped (Bi, Sb)$_2$Te$_3$ on Nb[23]) heterostructures. In such systems, the quality of the interface is vital in determining the key parameters, such as the superconducting coherence length.

On the other hand, if a single material can possess both superconductivity and topological surface states (TSSs), the construction of complicated heterostructures is unnecessary and the difficulty in achieving a perfect interface is avoided. Up to date, despite the large number of superconductor candidates examined, the observation of TSSs and/or MBSs/MFs have only been achieved in a few compounds (including iron-based superconductors FeTe$_{0.55}$Se$_{0.45}$[24,25,26,27], Li(Fe,Co)As[26], (Li$_{0.84}$Fe$_{0.16}$)OHFeSe[28], 2M-WS$_2$[29,30], PbTaSe$_2$[31,32,33] and β-Bi$_2$Pd[34,35,36]).

Recently, a layered transition-metal trichalcogenide superconductor, TaSe$_3$ (T$_c$ ~ 2K[37]), has been theoretically proposed to be a strong topological insulator with topological surface states (schematic illustrated in Fig. 1(a)), therefore becoming a candidate of TSC[38]. Compared to other TSC candidates, TaSe$_3$ is stoichiometric (compared to the FeTe$_{0.55}$Se$_{0.45}$), exfoliable (compared to the iron-based superconductors and β-Bi$_2$Pd), stable in ambient environment (compare to the 2M-WS$_2$) and with the TSS at the Fermi-level (compared to PbTaSe$_2$). In addition, in the monolayer limit[39], a quantum spin Hall insulator phase is expected so that the topological edge state is expected to form a 1D TSC (see Supplemental Experimental Procedures (SEP), part I and Ref. 40). Furthermore, TaSe$_3$ contains many other intriguing physical properties such as large non-saturating magnetoresistance[41,42], charge density wave in mesowires[43], high current-carrying capacity[44,45], and low levels of noise spectral density[46], making it an important platform to study the nature of TSCs as well as possible applications such as interconnects in quantum computers[44,45].

Despite the strong motivations, the experimental confirmation of topological electronic structure of TaSe$_3$ is challenging. Although there have been various indications, (e.g. the magnetic torque oscillation measurements have revealed the topological effects of the superconducting vortex state in a TaSe$_3$ ring crystal[47], and the Shubnikov–de Haas quantum oscillation reveals a non-trivial Berry phase[41], etc.), a direct observation of the topological surface states (TSSs) is lacking.

In this work, we report a comprehensive study on the electronic structure of TaSe$_3$. The high quality of the samples used in this work is confirmed by the scanning tunneling microscopy (STM) measurements which revealed clean cleaved surface along the $[10\bar{1}]$ direction and a persistent superconducting gap across quasi-1D TaSe$_3$ chains. By conducting complementary synchrotron and lab based angle-resolved photoemission spectroscopy (ARPES) measurements, we are able to identify both the inverted bulk band structure and the TSSs in the bulk bandgap. These observations, in nice agreement with the theoretical calculations and transport measurements, prove TaSe$_3$ as a superconductor with TSSs and thus an ideal platform for the investigation of topological superconductivity.

## II. RESULTS

### A. General Sample Characterizations

TaSe$_3$ crystalized into a monoclinic layered structure with space group $P2_1/m$ (No.11)[48]. The basic building blocks of TaSe$_3$ are two types of parallel trigonal-prismatic chains (labeled as red and blue in Fig. 1(b)) along the b axis, forming the quasi-1D crystal structure. Each chain is comprised of a linear stacking of irregular prismatic cages with six selenium atoms at the corners and one tantalum atom at the center. Each unit cell of the TaSe$_3$ consists of two type-I and two type-II chains related by inversion symmetry as shown in Fig. 1(b). Recent theoretical calculations indicate a topologically non-trivial band inversion occurs around B point (Fig. 1(c))[38] in the Brillouin Zone (BZ, detailed illustration can be found in Fig. 3(a)), due to the effect of strong spin-orbit interaction, resulting in an inverted bandgap with in-gap TSSs.

The quality of the TaSe$_3$ single crystal for this work was characterized by the single-crystal X-ray diffraction (XRD) and the X-ray photoemission spectroscopy (XPS) (Fig. 1(d)). The XRD patterns confirmed the crystal structure without impurity phases and the XPS spectrum shows sharp characteristic core-level peaks from both Ta and Se atoms, indicating the high quality of the crystals.

### B. Superconductivity and Surface morphology

The electronic properties of the crystal were firstly characterized by transport measurement. The temperature-dependent longitudinal resistance R$_{xx}$ along the prismatic chains indicates a superconducting phase transition at T$_c$ = 2K with a transition width of 0.5K (Fig. 2(a)). With the external magnetic field perpendicular to the sample surface, the superconductivity could be suppressed. It is worth noting that a finite resistance residue was observed below transition temperature, in line with the previous reports[49,50]. This is possibly due to the weakly coupled superconducting segments separated by non-superconducting regions caused by impurities[51].

To obtain the carrier concentration, magneto-transport measurements were carried out under a high magnetic field up to 12 T, with the temperature ranging from 1.5 K to 9 K (Fig. 2(b)). Non-saturating large magneto-resistance (inset of Fig. 2(b)) was observed with clear Shubnikov–de Haas (SdH) oscillations, confirming the high quality of the sample. Two main peaks with frequencies of F$_1$=84 ± 8 T and F$_2$=180 ± 9 T were deduced by fast Fourier transformation (See SEP, Part II), indicating the existence of two FS pockets with area of 0.008 ± 0.001 Å$^{-2}$ and 0.017 ± 0.001 Å$^{-2}$, in consistency with recent reports[41,42].

Since the prismatic Ta-Se chains have strong bonding along the *a* and *b* axis but relatively weak bonding along the *c* axis, the compound is easily cleaved along the $[10\bar{1}]$ direction, as previously been illustrated in Fig. 1(b). The morphology of the cleaved surface was investigated by STM measurement. As shown in Fig. 2(c), clear atom arrangement was revealed with the Ta-Se chains forming quasi-1D structures along the [010] direction (b axis). The distances between the atoms match well with the crystal lattice.

To confirm the superconducting behaviour of the compound, scanning tunneling spectroscopy (STS or dI/dV spectrum) measurements were performed at 1.2K right below $T_c$ and a typical spectrum is shown in Fig. 2(d). The dip on the Fermi edge indicates the formation of a superconducting gap, which could be well fitted by Bardeen–Cooper–Schrieffer (BCS) theory (detailed fitting method can be found in SEP, Part III). The appearance of the superconductivity gap in an area was further confirmed by STS spectra at several consecutive locations crossing the Ta-Se chains, as illustrated in Fig. 2(e). The clean surface atom arrangement as well as the well-defined persistent superconducting gap suggests $TaSe_3$ a suitable platform for further research on its novel quantum phenomena, e.g. Majorana Zero Modes.

### C. Overall Electronic Structure

The overall electronic structure of $TaSe_3$ was obtained by synchrotron based ARPES. After a set of photon energy dependent measurements to identify the $k_z$ locations (details can be found in the SEP, part IV), we chose the 30eV photons so that the measurement covers the B point in the 3D BZ. Due to strong $k_z$ broadening effect at low photon energy, we plot the data in the projected BZ on the $(10\bar{1})$ surface and compare the experiment results to the calculated bulk bands with integrated $k_z$ values (the relation between bulk and surface BZ is indicated in Fig. 3(a)). From the 3D plot of the electronic structure (Fig. 3(b)), constant energy contours at different binding energies (Fig. 3(c)) and the band dispersion along high symmetry directions ($\widetilde{Y} - \widetilde{\Gamma} - \widetilde{Y}$ and $\widetilde{S} - \widetilde{X} - \widetilde{S}$ directions, see Fig. 3(d), also SEP, Part V), we observe the following: 1. The shape of the Fermi surface (FS) shows clear anisotropy, consistent with the quasi-1D crystal structure of TaSe$_3$. 2. An electron pocket centered at the $\widetilde{X}$ point (labeled as "α") and a hole pocket centered at the $\widetilde{\Gamma}$ point (labeled as "β") can be identified in the FS, proving the overall semimetal nature of the compound, consistent with magneto-transport result. 3. The measured electronic structure shows nice agreement with the surface state calculation with Wannier functions. Specifically, at the $\widetilde{X}$ point (the projection of B point), where the band inversion occurs, we notice the α band appears to intersect with a hole-like band (labeled as "γ") thus forms the inverted bandgap and in-gap TSSs as predicted in the calculations[42] (illustrated by the zoom-in plot in Fig. 3(e)). However, due to the relatively low resolution of the synchrotron based ARPES system and the strong $k_z$ broadening effect, the details of the inverted band structure and the TSSs could not be clearly resolved, which motivated us to carry out the laser based ARPES measurements with much higher resolution as discussed below.

### D. Topological Surface States

To reveal the details of the inverted band structure and search for the TSSs, the state-of-art laser ARPES ($h\nu = 6.994\ eV$) measurement was performed with high energy, momentum, and spatial resolutions. The measured photoemission spectra are illustrated in Fig. 4. Due to the monoclinic crystal symmetry, the two $\widetilde{X}$ points in the projected surface BZ (labeled as $\widetilde{X}_1$ and $\widetilde{X}_2$) are inequivalent for the measured bulk electronic structure at certain $k_z$ value (the corresponding $k_z$ plane of the laser ARPES measures is indicated by the grey plane in Fig. 4(a)), which can be clearly seen from the FS map (Fig. 4(b)(i)) where the FS pockets from the α band were only observed around the $\widetilde{X}_1$ point while an additional pocket appears around both $\widetilde{X}$ points. This result was confirmed by using differently polarized lights, to minimize the uncertainty brought by the matrix element effect during the photoemission process[52] (see Fig. 4(b)(ii) for the result from linear vertical (LV) polarized photons, and the results from other polarizations can be found in SEP, Part VI). From these differences, we conclude that the α band corresponds to the bulk electron pocket and the additional features are the TSSs (detailed structure will be illustrated later), in agreement with the illustration in Fig. 4(a).

To compare our results with the magneto-transport measurement, we can estimate the FS pockets' area, as shown in Fig. 4(b)(ii), the area of the electron (from α band) and hole (from β band) pockets are estimated as $0.0248 \pm 0.005$ Å$^{-2}$ and $0.007 \pm 0.003$ Å$^{-2}$ (details in SEP, Part VII), respectively, which is consistent with the $180 \pm 9$ T (or $0.017 \pm 0.001$ Å$^{-2}$) and $84 \pm 8$ T (or $0.008 \pm 0.001$ Å$^{-2}$) frequency observed in the SdH quantum oscillations, and the small difference may be due to the band bending effect of the sample surface caused by surface potential.

We can further confirm the nature of the bulk and surface states by examining the dispersions along the high symmetry direction through both $\widetilde{X}$ points. As can be clearly seen in Fig. 4(c)(i), the α band shows two 'V' shape pockets along $\widetilde{S} - \widetilde{X}_1 - \widetilde{S}$ direction, while such pockets are missing along $\widetilde{S} - \widetilde{X}_2 - \widetilde{S}$ direction (see Fig. 4(c)(ii)). Similarly, the γ band shows a 'M' shape dispersion in Fig. 4(c)(i) but a 'Λ' shape dispersion in Fig. 4(c)(ii). The different shape of the dispersions clearly demonstrates the bulk band nature of the α and γ bands, which form a band inversion with a local bandgap (~20 meV, indicated in Fig. 4(c)(i), detailed analysis can be found in SEP, Part VIII).

Besides the bulk α and γ bands, the TSS can also be seen in Fig. 4(c) (marked as red dashed lines), in nice agreement with the calculations[42]. The constant energy contours showing the shape and evolution of the TSSs are also plotted with different binding energies in Fig. 4(d).

To illustrate the details of the TSSs (the schematic is shown in Fig. 4(e) according to the calculations), we further measured the band dispersions along the $\tilde{S} - \tilde{X}_2 - \tilde{S}$ and $\tilde{\Gamma} - \tilde{X}_2 - \tilde{\Gamma}$ directions at an elevated temperature in order to reveal the details slightly above the $E_F$ by thermal excitation. Here the spectra cutting across point $\bar{X}_2$ were chosen to avoid interference from the bulk α band as discussed above. From the spectra (Fig. 4(f)) along both directions, the TSSs are more dispersive along $\tilde{S} - \tilde{X} - \tilde{S}$ direction. On the other hand, the TSSs forms flatter bands along the $\tilde{\Gamma} - \tilde{X} - \tilde{\Gamma}$ direction. The energy position and shape of the TSS show nice agreement with the slab calculations (Fig. 4(f)(iii, vi)).

### III. DISCUSSION

The topological electronic structure and the superconductivity in TaSe$_3$ could be tuned relatively easily. As an example, we found the in-situ potassium dosing effectively increased the size of the electron pocket at $\tilde{X}$ (details in SEP, Part IX), which may lead to the change of the pairing mechanism and enhance the superconducting temperature (see Ref. 53). In addition, it has been proposed[38] that TaSe$_3$ could easily go through a topological phase transition under strain. Finally, as a layered quasi-1D compound, TaSe$_3$ could be thinned down to low dimension (2D or even 1D) and show more interesting physical phenomena. For example, in its 2D limit, TaSe$_3$ is proposed to be a quantum spin Hall insulator (See SEP, Part I and Ref. 40) with 1D helical edge states. If the superconductivity remains, the 2D system could be a candidate of 1D topological superconductor.

In conclusion, taking advantage of the large photon energy range of the synchrotron based ARPES and the high resolution of laser ARPES, we successfully observed the TSSs in quasi-1D superconductor $TaSe_3$, proving it as a new candidate for TSCs with rich physics and application potentials. In addition, STM result suggests a clean surface atom arrangement and persistent superconductivity band gap. Compared with existing intrinsic TSC candidates[24,29], $TaSe_3$ is stoichiometric, exfoliable and stable in the ambient environment, making it a great platform for the study of novel phenomena in TSCs and device applications. On the other hand, the non-ideal crystal quality of $TaSe_3$ (e.g. the existence of finite residue resistance below $T_c$), as well as the large density of bulk states crossing the $E_F$, suggest that further improvement of the material quality is required, and the search for ideal topological superconductors should be continued.

## IV. EXPERIMENTAL PROCEDURES

### A. Resource Availability

**Lead Contact**

Yulin Chen, Email: yulin.chen@physics.ox.ac.uk

**Materials Availability**

This study did not produce new unique materials.

**Data and Code Availability**

All data needed to evaluate the conclusions of the paper are present in the paper and/or the

Supplemental information. Additional data related to this paper may be requested from the authors.

### B. Crystal Synthesis

High-quality single crystals of TaSe$_3$ were grown from a self-flux method as described in Ref [13]. Starting materials of Ta powder (99.9%, Aladdin) and Se (99.999%, Aladdin) granules were mixed in a molar ratio of 1 : 6 and placed into an alumina crucible. The assembly was heated in a furnace up to 750 $^o$C within 9 hrs and kept for 20 hrs, and then slowly cooled down to 450 $^o$C at a temperature decreasing rate of 1.5 $^o$C/h. The excess Se was removed at this temperature by quickly placing the assembly into a high-speed centrifuge. The quartz tube was subsequently cooled to room temperature in air. The shiny quasi-1D crystals with a typical dimension of 2.4 × 0.1 × 0.06 mm$^3$ were obtained.

### C. Single-crystal X-ray Diffraction

Single-crystal XRD was performed using Mo target by the Rigaku Oxford Diffraction at the Department of Physics, University of Oxford. Measurement was done at room temperature. The beam spot size was 10-200 microns in diameter. The data was collected and analyzed by the CrysAlisPro software.

### D. Transport measurements

The electrode contacts were made by DuPont silver paste. Low-temperature transport measurements were employed in a cryofree superconducting magnet system (Oxford Instruments TeslatronPT) with base temperature of 1.47 K (down to 0.5K using helium 3) and maximum magnetic field of 12 T. The magnetic field was always applied perpendicular to the sample surface, along $[10\bar{1}]$ direction of the crystal. The sample resistance in four-terminal configuration was determined by exploiting standard lock-in techniques with an excitation current of ~2 μA at 13 Hz, and the current is applied along the quasi one-dimensional chain of TaSe$_3$. The uncertainty of the oscillation frequency is estimated by the FWHM of the FFT peaks.

### E.  Scanning tunnelling Microscopy

The STM/STS experiments performed at ShanghaiTech University. TaSe$_3$ single crystals cleaved inside ultra high vacuum at room temperature. Cleaved samples were transferred to a cryogenic stage kept at 77 K and 1.2K for STM/STS experiments. PtIr tips were used for both imaging and tunneling spectroscopy. The tips were calibrated with the surface states of silver islands on Si(111)–7 × 7. Lock-in technique was used to obtain superconducting gap dI/dV curves. A 0.1mV modulation signal at 997.233 Hz was applied to the sample together with the DC sample bias.

### F.  Angle-Resolved Photoemission Spectroscopy

Synchrotron based ARPES measurements were performed at beamline BL5-2 of Stanford Synchrotron Radiation Laboratory (SSRL), SLAC, USA. The samples were cleaved *in situ* and measured under ultra-high vacuum below 3×10$^{-11}$ Torr. Data were collected by a DA30L analyzer. The total energy and angle resolutions were 10 meV and 0.2°, respectively.

High spatial resolution laser-based ( $hv = 6.994\ eV$ ) ARPES measurements were performed at ShanghaiTech University. The samples were cleaved in situ and measured under ultra-high vacuum below 3×10$^{-11}$ Torr. Data were collected by a DA30L analyzer. The total energy and angle resolutions were 2.5 meV and 0.2°, respectively.

### G.  First Principle Calculations

The first-principles calculations were carried out by using the full-potential linearized augmented planewave method implemented in the WIEN2K simulation package[54]. The exchange and correlation potential was treated by the modified Becke-Johnson functional[55]. SOC was considered self-consistently. The radii of the muffin-tin sphere $R_{MT}$ were 2.5 bohrs and 2.38 bohrs for Ta and Se, respectively. The sampling of the BZ in the self-consistent process was taken as the grid of 7 × 19 × 6. The truncation of the modulus of the reciprocal lattice vector $K_{max}$ was set to $R_{MT}*K_{max} = 7$. The $d$ orbitals of Ta and the $p$ orbitals of Se were used to construct the maximally Wannier functions[56], which were then used to calculate the surface states. The slab calculation in Fig. 4(f) used a model of 40 layers.

## ACKNOWLEDGEMENTS

The work is supported by the National Key R&D program of China (Grants No.2017YFA0305400, No.2018YFA0307000, and No. 2017YFA0304600), the National Natural Science Foundation of China (Grants No. 11774190, No. 11874022, and No. 11674229), the Strategic Priority Research Program of Chinese Academy of Sciences (Grant No. XDA18010000) and EPSRC Platform Grant (Grant No. EP/M020517/1). Use of the Stanford Synchrotron Radiation Light Source, SLAC National Accelerator Laboratory, is supported by the US Department of Energy, Office of Science, Office of Basic Energy Sciences under Contract No. DE-AC02-76SF00515. This research used resources of the Advanced Light Source, a US DOE Office of Science User Facility under Contract No. DE-AC02-05CH11231. We thank the Elettra Light Source for access to the Spectromicroscopy beamline. H. T. Y. would like to acknowledge the support by the National Natural Science Foundation of China (51861145201, 91750101, 21733001). X. F. K. acknowledges the support from the National Natural Science Foundation of China (Grant No. 61874172). Z. J. W. acknowledges the support from the National Natural Science Foundation of China (Grant No. 11974395), the Strategic Priority Research Program of Chinese Academy of Sciences (Grant No. XDB33000000) and the CAS Pioneer Hundred Talents Program. D. P. and Y. W. L. acknowledge the support from Chinese Scholarship Council. L. X. Y. acknowledges the support from Tsinghua University Initiative Scientific Research Program.

## AUTHOR CONTRIBUTIONS

Y.L.C and Z.K.L. conceived the project. C.C. and A.J.L. performed ARPES experiment with the help of H.F.Y, Y.W.L., D.P. and L.X.Y.. D.H.L., M.H., A.B., C.J., A.B. and E.R.

**DECLARE OF INTEREST**

The authors declare no competing interests.

**FIGURE LEGENDS**

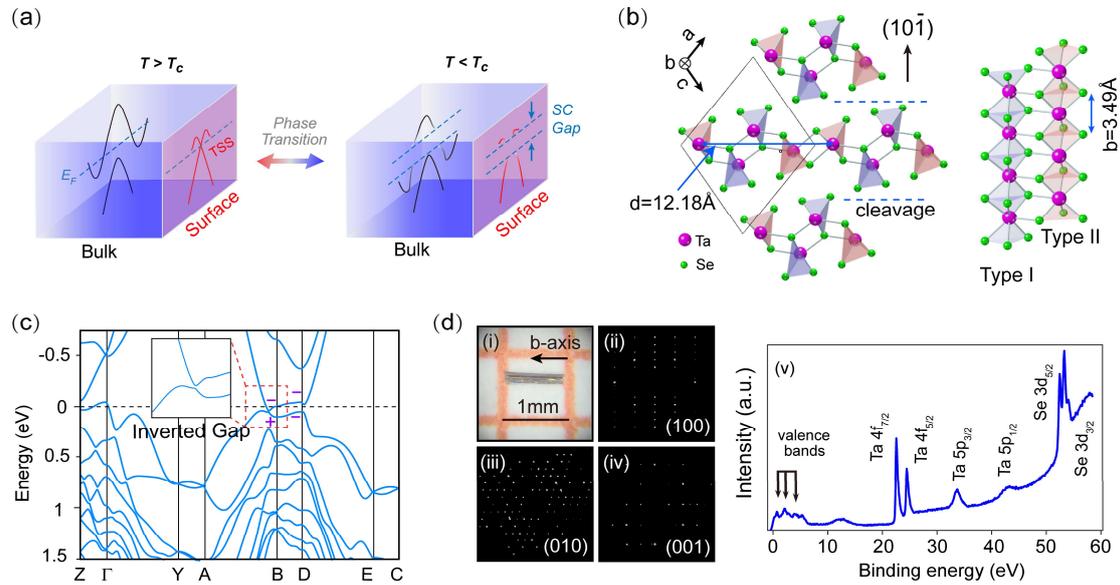

**FIG 1| General information of TaSe$_3$. (a),** Schematic of Topological superconductivity in TaSe$_3$. Above the superconducting transition temperature (T$_c$), the compound is a topological insulator with topological surface states (TSS) within the inverted band gap. Below T$_c$, both the bulk and surface states form superconducting gap. The bulk enters trivial superconducting state while the surface enters nontrivial topological superconducting state. **(b),** Monoclinic crystal structure of TaSe$_3$, formed by two types of Ta-Se chains (illustrated by different colors in the right panel) along the b axis. Cleavage planes are along the [10$\bar{1}$] direction and labeled by the blue dotted lines. **(c),** *Ab initio* calculation of the band structure, showing a nontrivial band inversion around B point which forms a strong topological insulating state (inset). **(d),** (i) Image of TaSe$_3$ single crystal. (ii-iv) X-ray diffraction patterns along (100), (010) and (001) directions and (v) corelevel photoemission spectrum showing the Ta *4f, 5p* and Se *3d* characteristic core levels.

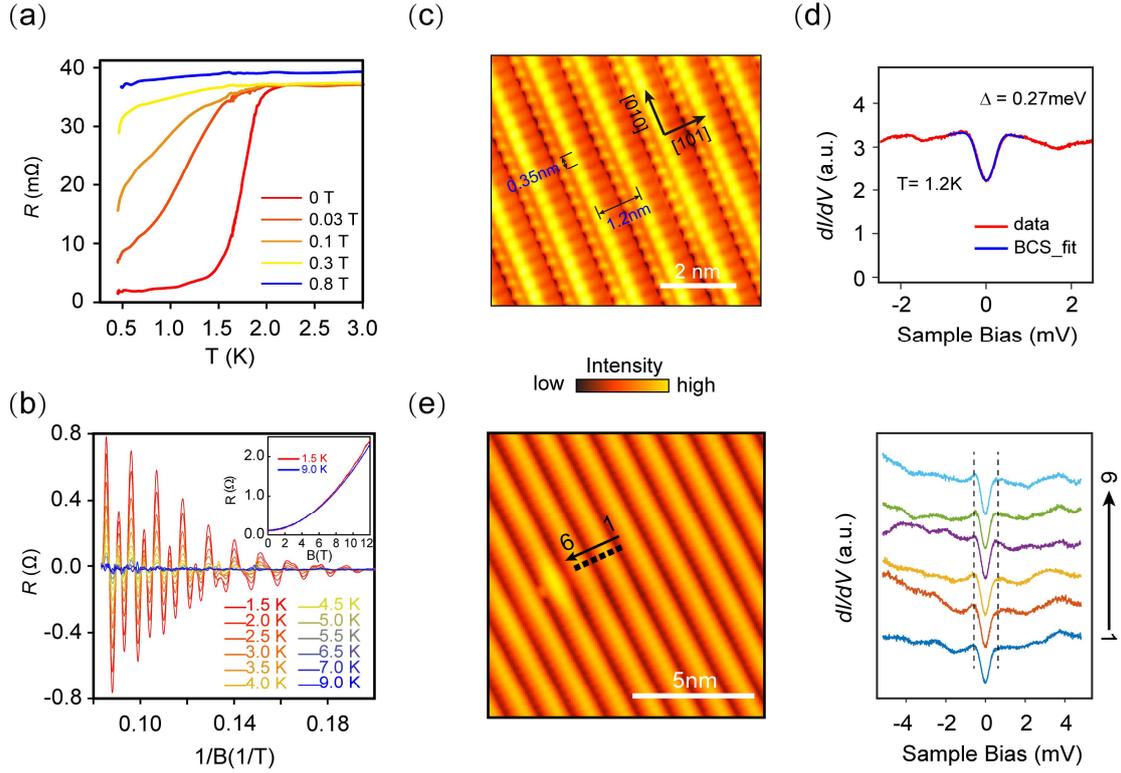

**FIG 2| Superconductivity and surface morphology. (a),** Resistance-Temperature curves under different magnetic fields perpendicular to the current direction, indicating a superconducting phase transition at ~ 2 K. **(b),** Shubnikov–de Haas (SdH) oscillatory component as a function of 1/B after subtracting the smooth background of the magnetoresistance at various temperatures. The inset plots the original magnetoresistance data measured at 1.5 K and 9 K. **(c),** STM image on the cleaved $(10\bar{1})$ surface, showing clean atoms arrangement, with sample bias $U_s$ =0.4 V, tunnelling current $I_s$ =250 pA, temperature T = 78 K. **(d)** Low energy dI/dV spectrum (set point: $V_s$=5 mV, $I_t$=200 pA) around Fermi edge, indicating the formation of the superconducting gap at 1.2 K. Blue line is the fitting result using a revised Dynes equation $\rho(E,\Gamma) = (E - i\Gamma)/[(E - i\Gamma)^2 - \Delta^2]^{1/2} + Slope \cdot E$. **(e),** dI/dV spectra (right panel, $U_s$=12 mV, $I_t$=200 pA, T=1.2 K) taken at 6 consecutive locations along the line indicated on the left image, showing the appearance of the superconducting gap cross the chains.

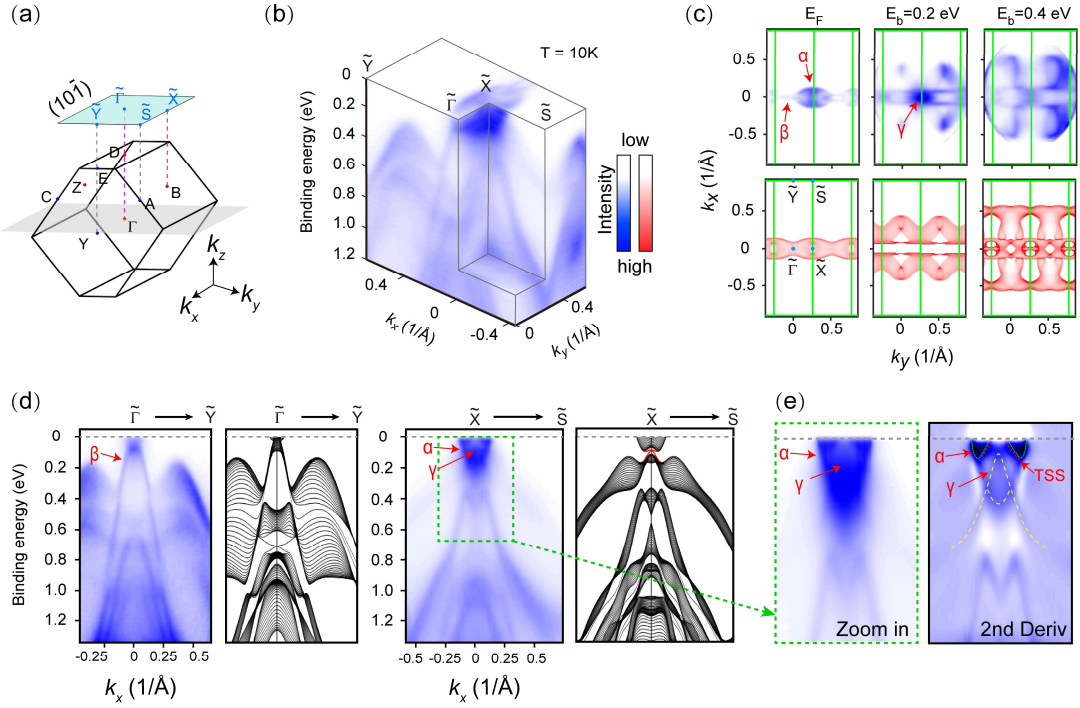

**FIG 3| Overall electronic structure. (a),** 3D Brillouin Zone (BZ) of TaSe$_3$ and the projected surface BZ on the $(10\bar{1})$ surface. High symmetry points are labeled. **(b),** 3D plot of the photoemission spectra on the cleaved $(10\bar{1})$ surface, labeled with the surface BZ. **(c),** Top row: Constant energy contours of photoemission spectra at different binding energies. Bottom row: Calculations at the same binding energies as on the top row. The green lines indicate the surface BZ. The bulk conduction band is labelled as α, while the bulk valance band at $\tilde{\Gamma}$ and $\tilde{X}$ are labeled as β and γ respectively. **(d),** Band dispersions along $\tilde{Y}-\tilde{\Gamma}-\tilde{Y}$ and $\tilde{S}-\tilde{X}-\tilde{S}$ directions, respectively, agree well with the related calculation. **(e)** Zoomed in plot (left panel) showing details of dispersions around $\tilde{X}$ point (marked by the green dashed rectangle in (d), and its 2nd derivative plot to illustrate the fine structure. Blue, yellow and red dashed lines are the guides of the eyes for the bulk conduction band (α), bulk valance band (γ) and the topological surface states, respectively. The data were collected using photons with hν=30 eV at 10 K.

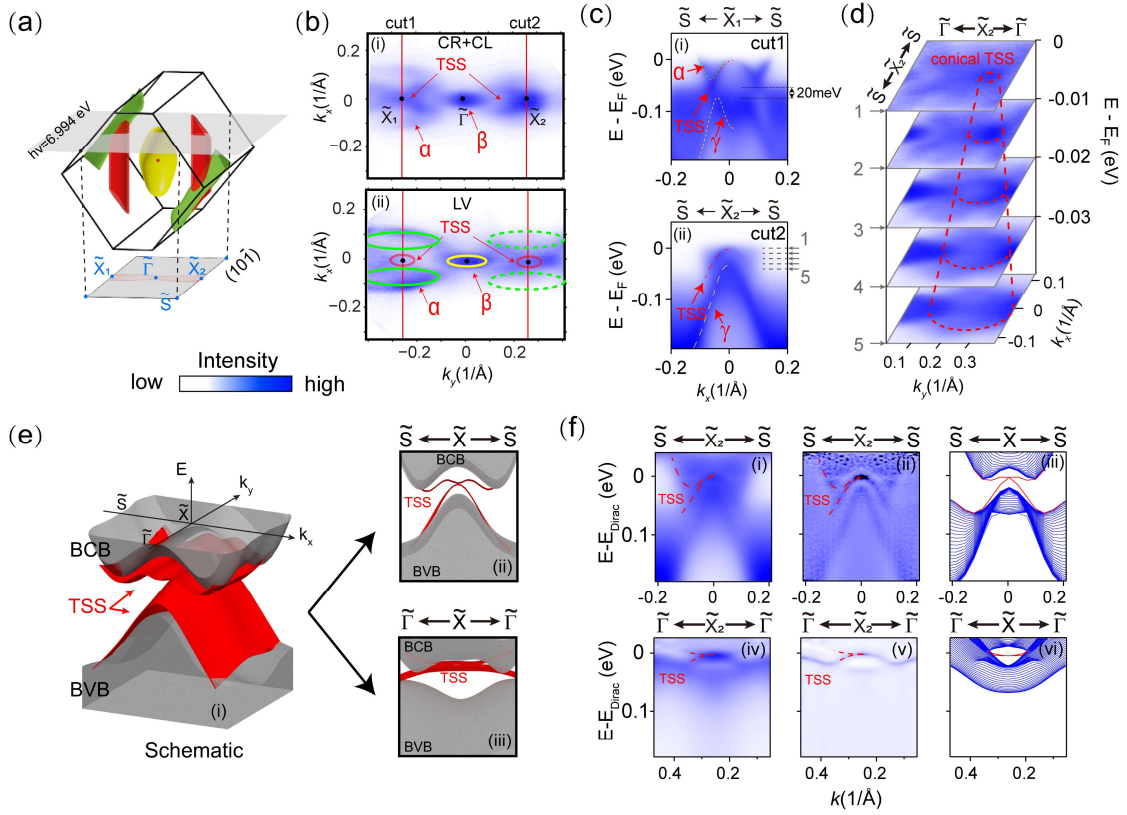

**FIG 4| Topological surface states. (a),** Schematic of 3D BZ and Fermi surface of TaSe$_3$. Green: bulk electron pockets (from α band). Yellow: bulk hole pocket (from β band). Red: surface electron pockets from the topological surface state (TSS). $\widetilde{X}_1$ and $\widetilde{X}_2$ denote the projection of two inequivalent $\widetilde{X}$ points on the $(10\bar{1})$ surface. Top grey plane indicates the $k_z$ location of the ARPES measurements in **(b, c, d, f)** with 6.994 eV photon. **(b),** (i) FS from the sum spectra of circular left (CL) and circular right (CR) photons. (ii) FS from the spectra of linear vertical (LV) photons. Different FSs are labeled by solid contours. The dashed green contours indicate the absent α bands' pockets adjacent to the $\widetilde{X}_2$ point. **(c),** Band dispersions along (i) $\widetilde{S} - \widetilde{X}_1 - \widetilde{S}$ and (ii) $\widetilde{S} - \widetilde{X}_2 - \widetilde{S}$ directions, with the guides to the eyes of the α, β and TSS bands. The inverted band gap (~20 meV) is also labeled in (i). **(d)** Constant energy contours at binding energies near E$_F$ (indicated by the grey arrows), where the evolution of the TSS is visualized by red dashed lines. **(e)** (i) 3D Schematic of the gapped bulk bands (grey) and the in-gap TSS (red). (ii, iii), 2D band dispersions along $\widetilde{S} - \widetilde{X} - \widetilde{S}$, $\widetilde{\Gamma} - \widetilde{X} - \widetilde{\Gamma}$ directions. BCB: bulk conduction band. BVB: bulk valence band. **(f)** (i) Raw data, (ii) maximum-gradient curvature plot[57], (iii) slab calculation of band dispersions along $\widetilde{S} - \widetilde{X}_2 - \widetilde{S}$ direction. (iv)-(vi) same as in (i)-(iii) for $\widetilde{\Gamma} - \widetilde{X}_2 - \widetilde{\Gamma}$ direction. Note that the data used in (i) and (iv) have been normalized along the k-axis to remove the effect of Fermi-Dirac distribution. Blue curves in (iii) and (vi) are the bulk band dispersion of different $k_z$, and the red curves denotes the TSSs. The measurement temperature is 80 K for **(b-d), (f)** (i-ii) and 30 K for **(f)**(iv-v).

**Supplemental Experimental Procedures**

**I. 2D quantum spin Hall insulator state in monolayer TaSe$_3$**

As TaSe$_3$ is an exfoliable layered compound, it could be thinned down to the monolayer limit. It would be desirable to investigate the electronic structure and the topological properties at the two-dimensional limit. The monolayer TaSe$_3$ consists of four prismatic chains (two type-I and two type-II) aligned parallel in one rectangular unit cell, with lattice constant a=3.49 Å and b=12.18 Å (Fig. S1(a)). The corresponding BZ is plotted in Fig. S1(a).

The *ab-initio* calculation result shows a much simpler band structure comparing with the bulk TaSe$_3$, where two electron pockets observed near X and a hole pocket near Γ, indicating a semimetal nature. The hole and electron pockets intersect near X and open an inversion gap, suggesting a quantum spin Hall insulator (QSHI) state (Fig. S1(b), even though a global gap doesn't exist). We further calculated the Z$_2$ invariant based on the parities of the occupied bands at the time-reversal-invariant momenta (TRIM) shown in Fig. S1(c). The Z$_2$ is computed to be 1, proving its topologically non-trivial nature.

Given the 2D QSHI nature, it is expected that one-dimensional helical edge states could be observed. The superconductivity in combination of the 1D helical edge states would allow us to propose a candidate for 1D topological superconductor which hosts Majorana Fermions near its boundary. Such possibility makes thin film TaSe$_3$ an important platform for the investigation of topological superconductivity.

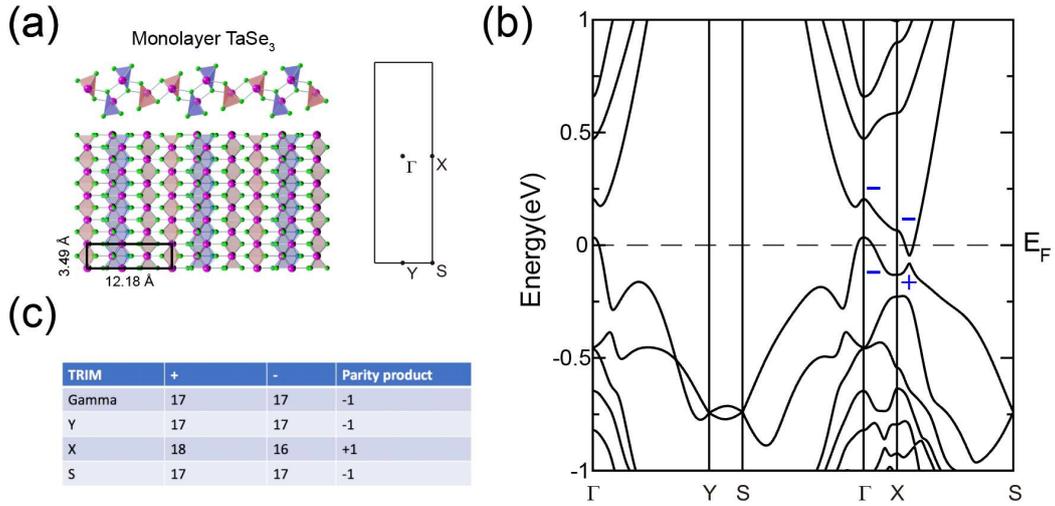

**FIG S1 | 2D quantum spin hall state in monolayer TaSe$_3$.** **(a)** Crystal structure of monolayer TaSe$_3$ (left panel) and corresponding BZ with high symmetry points labeled (right panel). **(b)** Calculated band structure of monolayer TaSe$_3$. **(c)** The parities of the occupied bands at the four TRIM.

## II. Transport measurement of TaSe3

Large range longitudinal resistance $R_{xx}$ measurement was shown in Fig. S2(a), illustrating a metallic behavior of the compound. The inset shows 4-probe sample measuring geometry. The result of Fast Fourier Transformation (FFT) analysis of Fig. 2(b) from the main text was shown in Fig. S2(b). Two major peaks with frequency of $F_1$=84 T and $F_2$=180 T are observed, indicating two electron pockets. More transport results could be found in recent works on TaSe$_3$ [1,2].

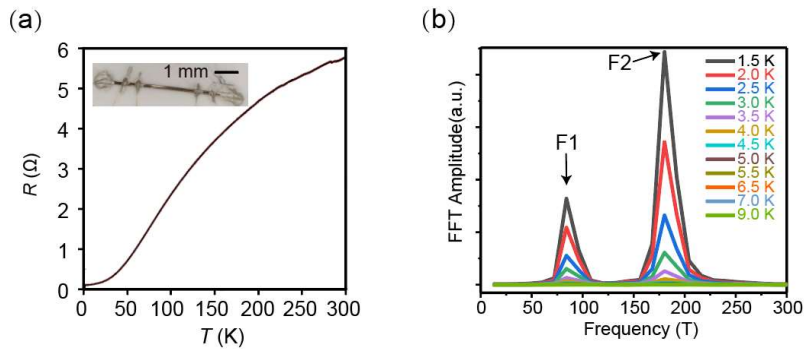

**FIG S2 | Transport Measurement of TaSe$_3$.** **(a)** Wide range Resistance-Temperature measurement of TaSe$_3$ crystal. Upper inset shows the sample used in the four-probe measurement. **(b)** FFT analysis of Fig. 2(b) in the main text.

### III. Superconductivity gap fitting of STS data

Conventional superconductivity can be well described by the BCS theory and the measured STS superconducting gap can be fitted by the Dynes equation:

$$\rho(E,\Gamma) = (E - i\Gamma)/[(E - i\Gamma)^2 - \Delta^2]^{1/2}$$

Where the $\Delta$ is the superconducting gap and $\Gamma$ is a broadening term, which is a function of the temperature. As the density of states measured by dI/dV curve is not constant near EF, it is legit to add a linear background term in the fitting function. Therefore the fitting function used here is:

$$\rho(E,\Gamma) = (E - i\Gamma)/[(E - i\Gamma)^2 - \Delta^2]^{1/2} + Slope \cdot E$$

The non-zero intensity at zero bias reflects the finite density of state due to the finite measurement temperature (~1.2K, only 60% of the Tc of TaSe$_3$). To better illustrate the result, a series of theoretical curves with different gamma values were plotted in Fig. S3 (a), together with the experiment result. We can find that the superconductive gaps become more broadened as $\Gamma$ increases (increasing temperature) and the intensity at zero bias increases. Given the small superconducting gap size, a zero intensity would only be achieved if the measuring temperature could be further reduced.

The fitting result of the STS spectra in Fig. 2(e) is plotted in Fig. S3(b). Small deviations were found among the curves probably due to the variation of local environment at each point.

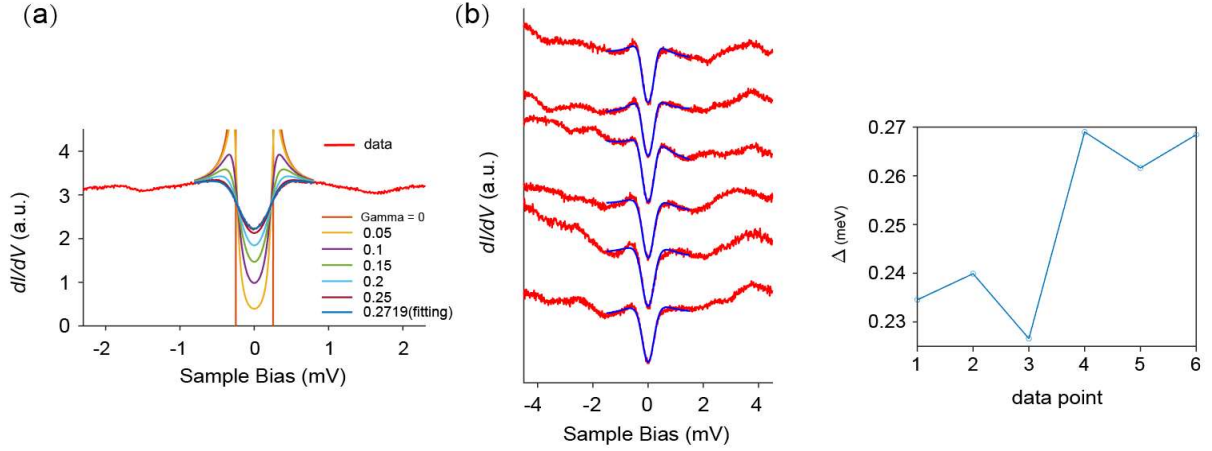

**FIG S3 | Fitting of STS data.** (**a**) Dynes equation with different gamma values (different temperature), indicating the broadening of the superconducting gap and the formation of non-zero intensity at zero. (**b**) Fitting of the measured superconductivity gap illustrated in Fig. 2(e) and the ∆ value.

### IV. Photon energy dependent measurement and $k_z$ evolution of the electronic structure of TaSe$_3$

During the photoemission process, the horizontal momentum ($k_{||}$) of band electrons are obtained directly from the free-electrons due to the momentum conservation law. The vertical momentum ($k_z$) is not conserved however. With the free-electron final state approximation and a potential parameter $V_0$ (also known as the inner potential) describing the energy difference of photoelectrons before and after passing the crystal surface, we can derive the $k_z$ as:

$$k_z = \frac{\sqrt{2m_e(E_k \cos^2\theta + V_0)}}{\hbar}$$

where $\theta$ is the emission angle and $E_k$ is the kinetic energy of the emitted electron, which satisfies:

$$E_k = h\nu - w - E_B$$

where $h\nu$ is the photon energy, $w$ is the work function of the sample and $E_B$ is the electron binding energy.

As $V_0$ varies with compounds, we typically performed energy dependent ARPES by using a broad range of photon energies to ensure that we can cover enough $k_z$–span (ideally more than one BZ), and use the high symmetry points in the $k_\parallel$-$k_z$ plane of the BZ to determine the $k_z$ value of each photoemission spectrum.

As for TaSe$_3$, due to its monoclinic crystal structure, the three-dimension BZ changes in size along the $\Gamma - X$ direction, making the determination of the k$_z$ period challenging (See Fig. S4(a)). Therefore, we carefully performed photon energy dependent ARPES measurement along the $\Gamma - Y$ direction (with k$_z$ independent length) by using photons from 26 eV to 40 eV.

From *ab-initio* calculation, the projected dispersions of the bulk bands along the $\tilde{Y} - \tilde{\Gamma} - \tilde{Y}$ direction which combines all the $k_z$ values were plotted in Fig. S4(b)(i), whereas the dispersion along the $Y - \Gamma - Y/E - D - E$ directions (i.e., $k_z$=0 and $k_z$=π/c) is plotted in Fig. S4(b)(ii) and (iii), respectively. By comparing the measured spectra (Fig. S4(c)) to the calculated band structure, we find the dispersions observed all better agree with the projected dispersions (Fig. S4(b)(i)) rather than dispersion at a specific $k_z$ value. Such observation is due to the severe $k_z$ broadening effect, which is the combined effect of the very small $k_z$ period of the BZ (~0.4 Å$^{-1}$ from $\Gamma - D$), as well as the insufficient $k_z$ resolution of low energy photons.

However, we could still estimate the $k_z$ value of each dispersion by tracing features that evolve rapidly with $k_z$. For example, the ε band labeled in Fig. S4(b)(ii)-(iii) is closest to the E$_F$ at $k_z$=0 while further away from E$_F$ at $k_z$=π/c. The same trend could be observed in the photoemission spectra. For spectra measured by photon energies ~34 eV, the intensity of the ε band accumulates near the top of the envelope. With higher or lower photon energies, the spectral weight is transferred towards higher binding energy, suggesting the downward shift of the ε band. Such evolution could be better visualized by plotting the evolution of the energy distribution curve at $k_x$=0.4 Å$^{-1}$ as a function of photon energy, where the clear periodic pattern could be observed (Fig. S4(d)(i)). Similar evolution could be found for the electron pocket (labeled as α band in Fig. S4(b)), which is below E$_F$ at D and above E$_F$ at Γ. For spectra measured by photon energies ~28 eV and ~40 eV, there is a strong intensity of the α band below E$_F$ and the spectra become electron like. For spectra measured by photon energy ~34 eV, the intensity of the α band is weakened and the hole pocket becomes obvious. Such evolution could also be visualized by plotting the evolution of the energy distribution curve at $k_x$ =0 as a function of photon energy, where the clear periodic pattern could be observed (Fig. S4(d)(ii)). Combining these pieces of information together, we could reliably determine the high symmetry points in the three-dimensional BZ and the $V_0$ is estimated to be 9eV. For example, the spectra measured with 28 eV and 40 eV photons are along the E − D − E direction ($k_z=\pi/c$) and the spectra measured with 34 eV photons are along the Y − Γ − Y direction ($k_z=0$). Therefore, the spectra analyzed in the manuscript are taken with 30 eV photons thus hold $k_z$ values between Γ and D, thus would have strong contribution from the B point.

While the synchrotron data could be interpreted with a $V_0$ as discussed above, it is hard to determine the exact $k_z$ value for the laser ARPES data. The reason is that when the excitation energy is low (e.g. 6.994 eV here), the final state of the excited photoelectron inside the material (before propagating to the material surface) could no longer be described by free electron, but a bloch-like states with relative flat dispersion. Therefore, the way to set a $V_0$, which was introduced to represents the pseudo-potential experienced by the electron from $E_F$ to a free-electron-like excited state is no longer valid. A detailed explanation can be found in the laser ARPES study on ZrTe$_5$ (Section C of Ref. 3).

With this consideration, the $k_z$ position of the laser data could only be estimated by the comparison with the theoretical calculation. From the analysis in Fig. S8, the bulk electron band (α) sits away from $E_F$ at the $\widetilde{X}_1$ point and close to $E_F$ at the $\widetilde{X}_2$ point. Together with the evolution of the valence band as indicated the yellow dashed line in Fig. 4(c)(i)-(ii), we find the band dispersions coincide with the two extremes in the calculation of the projected surface. Therefore, the $k_z$ position of the laser data is estimated to be around B point.

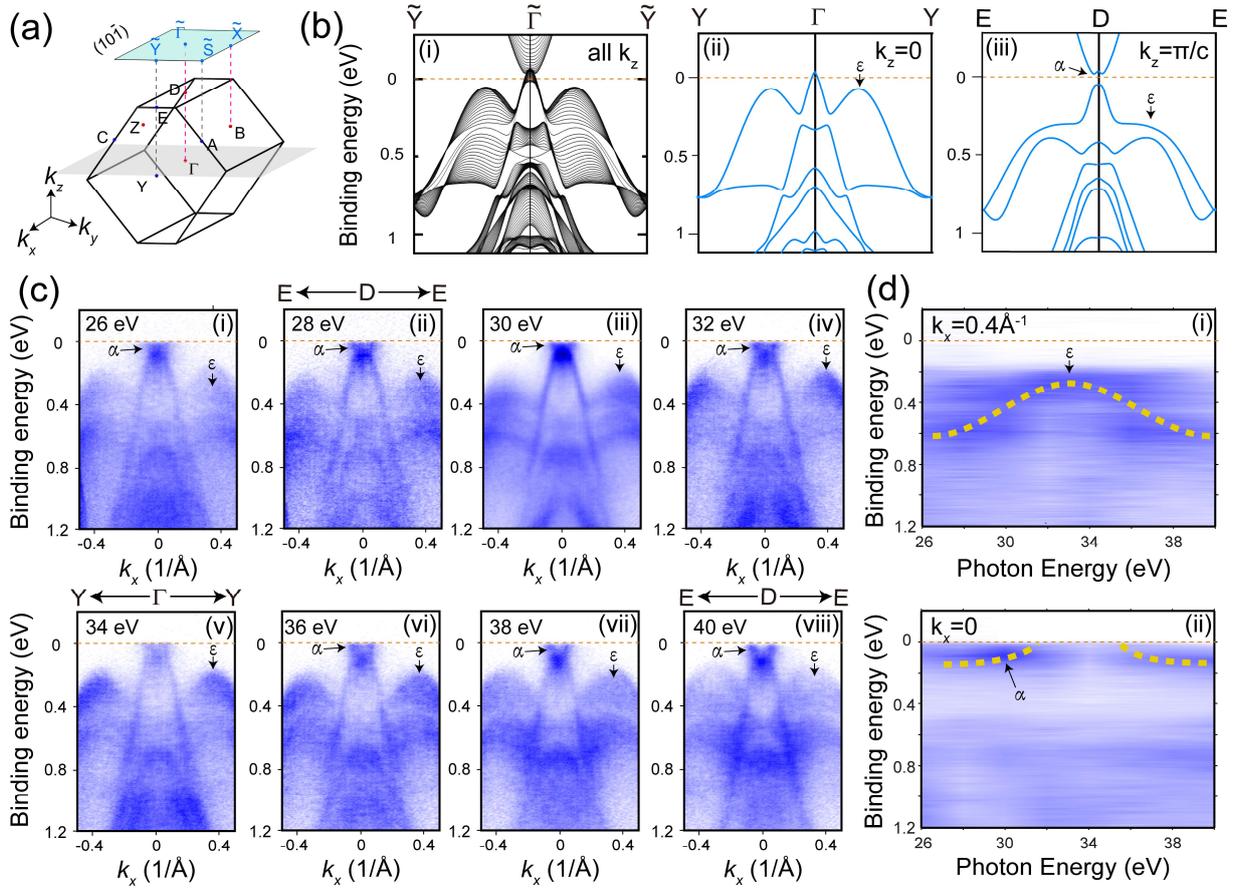

**FIG S4 | $k_z$ dependent measurement via continuously changing photon energy of TaSe$_3$.** (a) 3D Brillouin Zone of TaSe$_3$ and the projected surface Brillouin Zone onto the $(10\bar{1})$ surface. High symmetry points are labeled. (b) (i) Calculated bulk band structure along the $\tilde{Y} - \tilde{\Gamma} - \tilde{Y}$ direction. Bands with different $k_z$ values are all projected into the surface BZ. (ii, iii) Calculated bulk band structure along the $Y - \Gamma - Y$(ii)/$E - D - E$(iii). The α and ε bands are labeled for comparison. (c) The photoemission spectra of cuts along the $\tilde{Y} - \tilde{\Gamma} - \tilde{Y}$ direction measured with photon energies from 26 eV to 40 eV. The α and ε bands are labeled for comparison. (d). Intensity plot of the stacked energy distribution curves measured with different photon energy at $k_x=0.4$ Å$^{-1}$ (i) and $k_x=0$ (ii). Dispersions of the α and ε bands are labeled by the eye guides.

# V. Bulk band evolution on the projected $(10\bar{1})$ surface

As explained in the manuscript, due to the low crystal symmetry of TaSe$_3$, the measured electronic structure exhibits inequivalence between the $\widetilde{X}_1$ and $\widetilde{X}_2$ points and also adjacent $\widetilde{\Gamma}$ points. As illustrated in Fig. S5(a), a constant energy contour at certain $k_z$ values across two adjacent BZs would cut through two $\widetilde{\Gamma}$'s with different $k_z$ value (for example, the $\Gamma$ point in the first BZ and the D point in the second BZ). Therefore, we label the two adjacent $\widetilde{\Gamma}$ points as $\widetilde{\Gamma}_1$ and $\widetilde{\Gamma}_2$, respectively.

We present the constant energy map at the $E_F$ across two BZs using synchrotron ARPES with all the inequivalent high symmetry points labeled in Fig. S5(b). Evidently, we observed the different pocket sizes between the α band near $\widetilde{X}_1$ and $\widetilde{X}_2$ points, as well as the β band near $\widetilde{\Gamma}_1$ and $\widetilde{\Gamma}_2$ points. The comparison of the high symmetry cut along the $\widetilde{\Gamma}_2 - \widetilde{Y}/\widetilde{\Gamma}_1 - \widetilde{Y}$ and $\widetilde{X}_1 - \widetilde{S}/\widetilde{X}_2 - \widetilde{S}$ further elaborates the differences. While the projection of the α band is evident in the $\widetilde{\Gamma}_2 - \widetilde{Y}$ cut (Fig. S5(c)(i)), it is not observed in the $\widetilde{\Gamma}_1 - \widetilde{Y}$ (Fig. S5(c)(ii)). The β band top goes above $E_F$ near $\widetilde{\Gamma}_1$ but seems below $E_F$ near $\widetilde{\Gamma}_2$. The α band is clearly identified near $\widetilde{X}_1$ while becomes a faint feature with shrunk size near $\widetilde{X}_2$. All these observations are consistent with their different $k_z$ value in the 3D BZ and could be interpreted as the *ab-initio* calculation results (Fig. S5(c)(iii),(vi)) projected to a certain $k_z$ range. The evolution of the electronic structure across several BZs proves the bulk origin of the electron-like α and hole-like β bands.

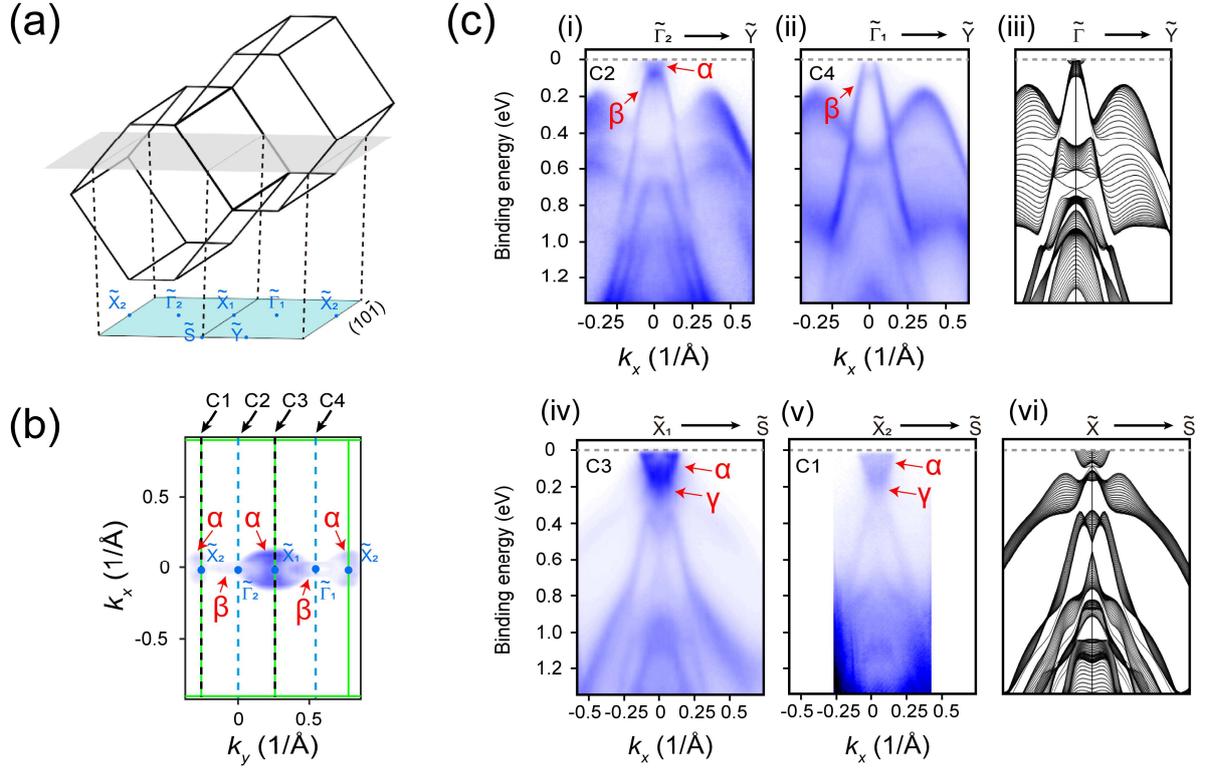

**FIG S5 | Bulk band evolution on the projected $(10\bar{1})$ surface across two BZs. (a)** Plot of two adjacent 3D BZs and their projection onto the $(10\bar{1})$ surface with high symmetry points labeled. Grey plane indicates the position of a constant energy contour across the two BZs. **(b)** Constant energy contour at $E_F$ measured with hv=30 eV photon across more than two BZs. High symmetry points and identified FSs in the surface BZ are labeled. C1-C4 labels the direction of the four high-symmetry cuts in **(c)**. **(c)** High symmetry cuts along the $\tilde{Y} - \tilde{\Gamma}_2 - \tilde{Y}$ (i), $\tilde{Y} - \tilde{\Gamma}_1 - \tilde{Y}$ (ii), $\tilde{S} - \tilde{X}_1 - \tilde{S}$ (iii) and $\tilde{S} - \tilde{X}_2 - \tilde{S}$ (iv) directions with several identified bands labeled. The results are compared with the *ab-initio* calculations results along the $\tilde{Y} - \tilde{\Gamma} - \tilde{Y}$ (iii) and $\tilde{S} - \tilde{X} - \tilde{S}$ (vi) directions.

## VI. Polarization dependent measurement of TaSe$_3$

We carried out polarization dependent measurements with the $hv = 6.994\ eV$ laser. The measurement geometry was illustrated in Fig. S6(a). Linearly vertically (LV) , linearly horizontally (LH), circularly left

(CL) and circularly right (CR) polarized laser beam were used for the measurement. We note due to the angle of beam incidence, the LV polarization is purely parallel to the sample surface while other polarizations consist of polarization components parallel to the normal of the sample surface.

The measured photoemission spectra using LV, CL, and CR photons were illustrated in Fig. S6(b-d), including the constant energy contours at $E_F$ (Fig. S6(b-d)(i)), high symmetry cuts along the $\tilde{S} - \tilde{X}_1 - \tilde{S}$ (Fig. S6(b-d)(ii)) and $\tilde{S} - \tilde{X}_2 - \tilde{S}$ (Fig. S6(b-d)(iii)) directions. All the bulk originated bands (α, β, and γ) and the TSS could be visualized with different parts being enhanced or suppressed due to the photoemission matrix element. We note the α band could be clearly visualized near $\tilde{X}_1$ in all three polarizations but disappears near $\tilde{X}_2$, the γ band appears to be a 'M' shape near $\tilde{X}_1$ but 'Λ' shape near $\tilde{X}_2$, thus proving the scheme due to projection from different $k_z$ values, as illustrated in Fig. 4(a) in the main text.

The other observation is that the hollow top of the 'M' shape γ band and the α band observed in the $\tilde{S} - \tilde{X}_1 - \tilde{S}$ cut shows synchronized enhancedment and suppression when probed with differently polarized photons (Fig. S6(b-d)(ii)). And the the hollow top shows different enhancement/suppression comparing with the rest of the γ band. These facts are consistent with the α-γ band inversion picture, suggesting that the hollow shape of the 'M' shape was the bottom of the electron-like α band before the band inversion taking place.

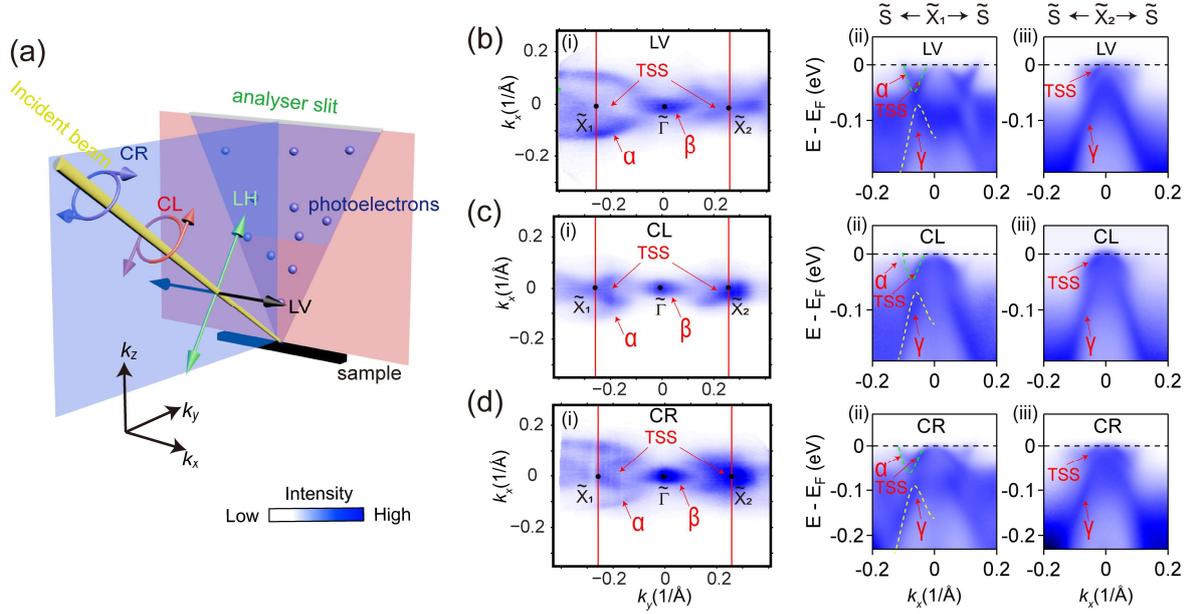

**FIG S6 | Polarization dependent measurement of TaSe$_3$. (a)** Measurement geometry of our laser ARPES system. LH: linear horizontal polarization. LV: linear vertical polarization. CL: circularly left polarization. CR: circularly right polarization. **(b)-(d)** Constant energy contours at E$_F$ (i), band dispersions along $\tilde{S}-\tilde{X}_1-\tilde{S}$ (ii) and $\tilde{S}-\tilde{X}_2-\tilde{S}$ (iii) directions measured with LV **(b)**, CL **(c)** and CR **(d)** photons. TSS and bulk bands (α, β and γ) are labeled.

## VII. Estimation of the Bulk Fermi Surface Area in TaSe$_3$

As discussed in the main text, the bulk band FS of TaSe$_3$ consists of two parts. One is the four electron-like α pockets near $\tilde{X}$, the other one is the hole like β pocket near $\tilde{\Gamma}$.

Due to matrix element effects of photoemission, we start from the sum spectra measured from LV and CL polarized photons, which has a large intensity of the whole contour of the α pockets, as shown in Fig. S7(a). The stacked plot of the energy distribution curves (EDCs) of Fig. S7(a) along $k_x$ are shown in Fig. S7(b), where clear curve peaks from the α pockets could be observed. Using multiple Lorentzians to fit each EDC (a typical fitting result is shown in Fig. S7(c); the fitted curve is the red curve marked in Fig. S7(b)), we traced the peaks of α pockets, which form the green dotted contour in Fig. S7(a). From here, we could reliably estimate the area of α pocket surrounded with in the contour by integration. The area of upper and lower two half α pockets are 0.0122 Å$^{-2}$ and 0.0126 Å$^{-2}$. A single full α pocket thus has an area of ~0.0248 ± 0.005Å$^{-2}$.

On the other hand, the contour of the β pocket near $\tilde{\Gamma}$ could not be reliably obtained by fitting with multiple Lorentzians due to the shape of the β band. The β band has the large effective mass along $k_x$ (Fig. S7(e)) and that the $E_F$ locates extremely close to the band top (Fig. S7(f)). Therefore, the single Lorentzian fitting would work better than multiple Lorentzian fitting along the $k_y$ direction. To avoid such ambiguity, we plot a slim elliptical shape to estimate the area of the β pocket, where the length of its long (along $\tilde{\Gamma}$-$\tilde{X}$) and short (along $\tilde{\Gamma}$-$\tilde{Y}$) axes could be determined by the spacing of the Fermi energy crossing points ($k_F$s) of the β band. By extracting the $k_F$ points of the β band along $\tilde{\Gamma}$-$\tilde{X}$ (Fig. S7(e)) and $\tilde{\Gamma}$-$\tilde{Y}$ (Fig. S7(f)) directions, the half long and short axes of the ellipse are estimated to be 0.085 Å$^{-1}$ and 0.025 Å$^{-1}$ and the area of the ellipse is 0.007 ±0.003 Å$^{-2}$, which can be viewed as a good estimate of the area of the hole-like β pocket.

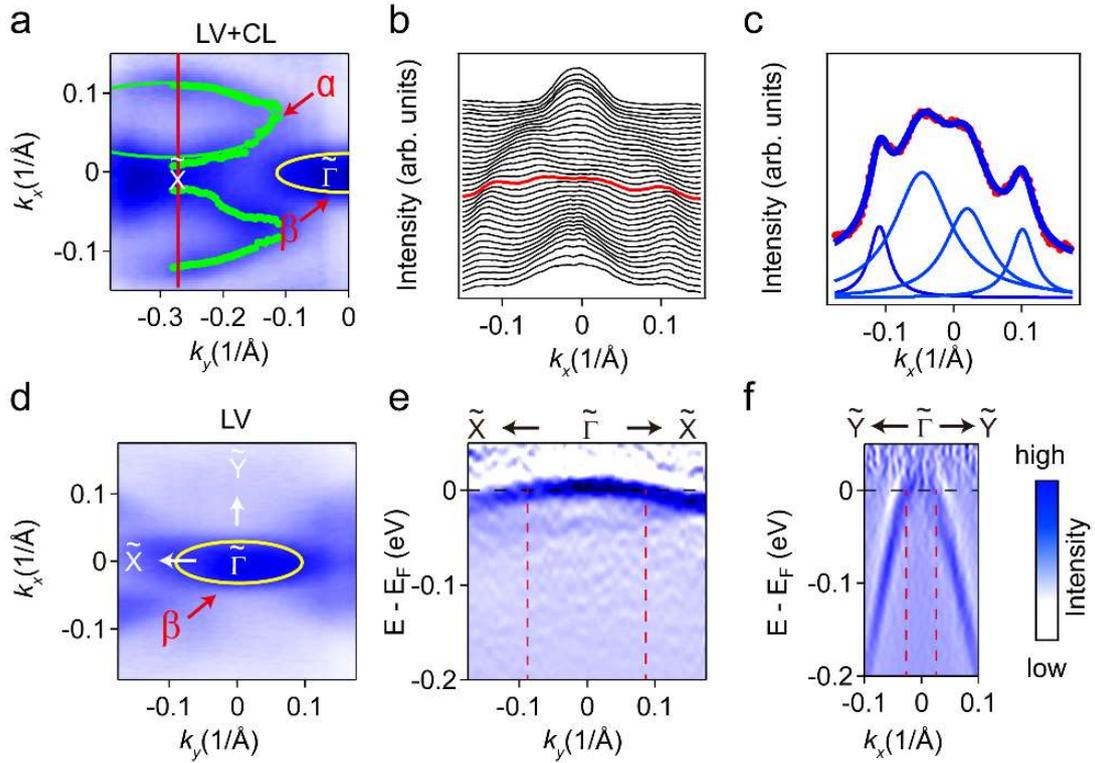

**FIG S7 | Estimate of the bulk Fermi surface area in TaSe$_3$. (a)** Fermi surface around zone boundary $\tilde{X}$. Sum spectra measured by LV and CL polarized photons are used to have a better visualization of the α pockets. The dotted green curves are generated by Lorentzian fitting of the contour intensity distribution as in (b). The solid green ellipse is shown to be a good estimate of the area of the α pocket. The red line shows the BZ boundary. **(b)** Stacked plots of the energy distribution curves (EDCs) along $k_x$ of Fermi surface in (a). The red curve is a typical EDC analyzed in (c). **(c)** Multiple Lorentzian fitting results of a typical EDC marked (b). **(d)** Fermi surface at zone center $\tilde{\Gamma}$ which is measured by LV polarized photons. The yellow ellipse shows the estimate of the area of the β pocket. **(e)** Band dispersion of the β band along the $\tilde{\Gamma}$-$\tilde{X}$ direction. The $k_F$s are labeled by two dashed lines ($\pm 0.085$ Å$^{-1}$). **(f)** Band dispersion of the β band along the $\tilde{\Gamma}$-$\tilde{Y}$ direction. The $k_F$s are labeled by two dashed lines ($\pm 0.025$ Å$^{-1}$). To better visualize of the β bands across the E$_F$, the phototoemission spectra has been devided by Fermi-Dirac distribution and the second derivative of the spectra are shown in (e-f).

## VIII. Detailed analysis of band disperions at X point

The original data in Fig. 4c(i) (along the $\tilde{S} - \tilde{X}_1 - \tilde{S}$ direction), its second derivative and EDCs are illustrated in Fig. S8 (left column). An asymmetric behavior of the upper electron bands (most likely due to the complex matrix element effects) is observed. There are two electron bands on the left branch of the electron bands, which are very close to each other but could still be separated (Fig. S8(b)). One of them is contributed by the bulk α band and the other should be from the upper branch of TSS. These two features would appear as one peak and one shoulder on the plot of the stacked EDCs shown in Fig. S8(c). On the other hand, the lower branch of the TSS is very hard to be separated from the α band here. However, the intensity going down and merged into the valence band could still be the hint of the lower branch of TSS.

The detailed band dispersions along $\tilde{S} - \tilde{X}_2 - \tilde{S}$ direction are plotted in Fig. S8 (right column). We could find that the TSS exists at exactly the same position while the α band could only be found at a higher energy position around Fermi level. This is consistent with that the bulk bands evolve with different $k_z$ while the surface states remain unchanged.

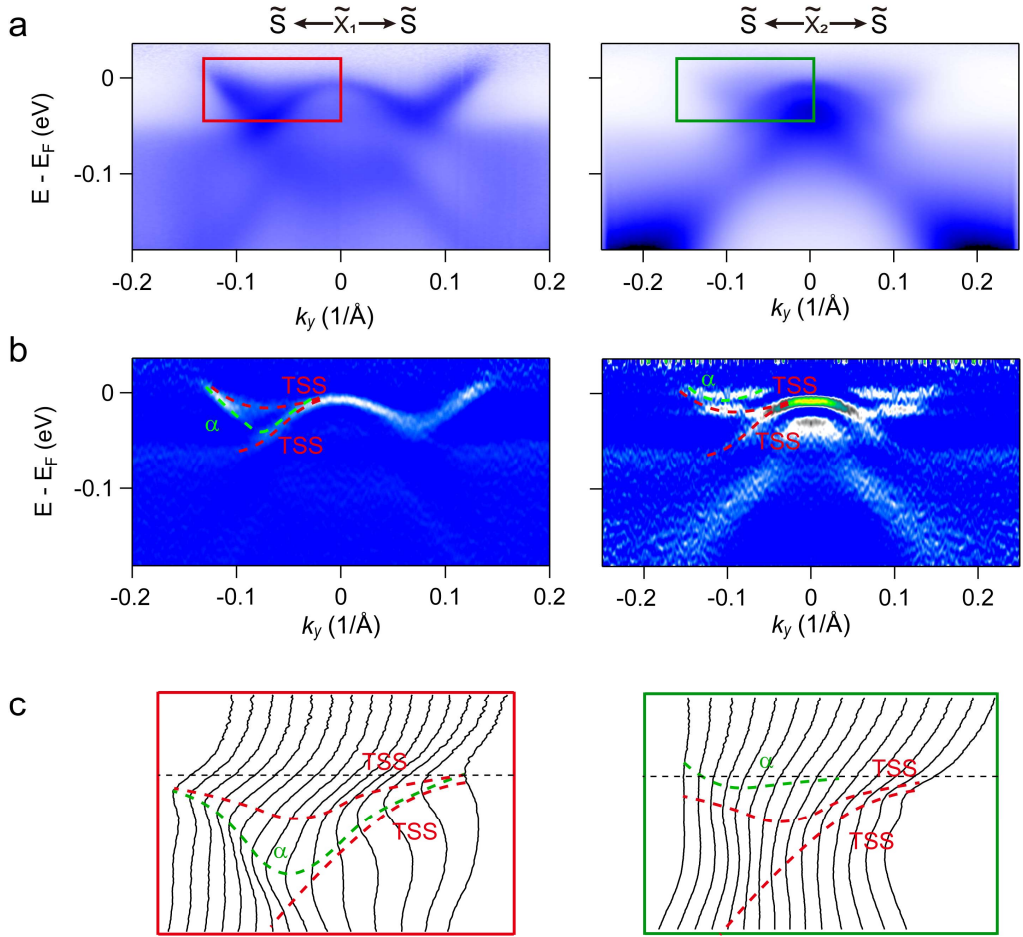

**FIG S8 | Detailed analysis of band dispersions.** ARPES data taken by LV photons along $\tilde{S} - \tilde{X}_1 - \tilde{S}$ and $\tilde{S} - \tilde{X}_2 - \tilde{S}$ directions (a) Original data. (b) second derivative image of (a) with respect to energy (y direction). (c) selected EDCs in from image marked by red (green) box in (a). TSS: topological surface states.

### IX. Tuning of the electronic structure of TaSe$_3$ by alkaline atoms dosage

The electronic structure of TaSe$_3$ could be modified by dosing the sample surface with alkaline metal (potassium). The SAES alkaline metal dispenser was used for the experiment. As Fig. S9 has shown, after potassium dosage, all the bulk bands were shifted towards the higher binding energies. The electron pocket at $\tilde{X}$ (the α band) shifted ~200 meV (see the intensity plot of the high symmetry cuts shown in Fig. S9(a)) and increases in the Fermi surface area (see the three-dimensional volume plot of the electronic structure in Fig. S9(b) and the constant energy contours in Fig. S9(c)). The electron pocket extends towards the $\tilde{\Gamma}$ point so that it could be observed in the dispersion along the $\tilde{Y} - \tilde{\Gamma} - \tilde{Y}$ direction (Fig. S9(b)). At the same time, the hole-bands (the labelled β and γ bands in Fig. S9) also shifts towards higher binding energies. The hole pocket on the Fermi surface becomes a spot like (see Fig. S9 (c)), indicated that the electron-type carrier dominates the transport after the potassium dosage. The band inversion of the α and γ bands still persists, suggesting the robustness of the topological electronic states. We note the measured constant energy contours (Fig. S9(c)(i)) shows the nice agreement with the calculation results of the pristine sample by shifting the Fermi energy 200 meV towards high kinetic energy (Fig. S9(c)(ii)), suggesting the effect of potassium dosage is a rigid shift of the electronic bands.

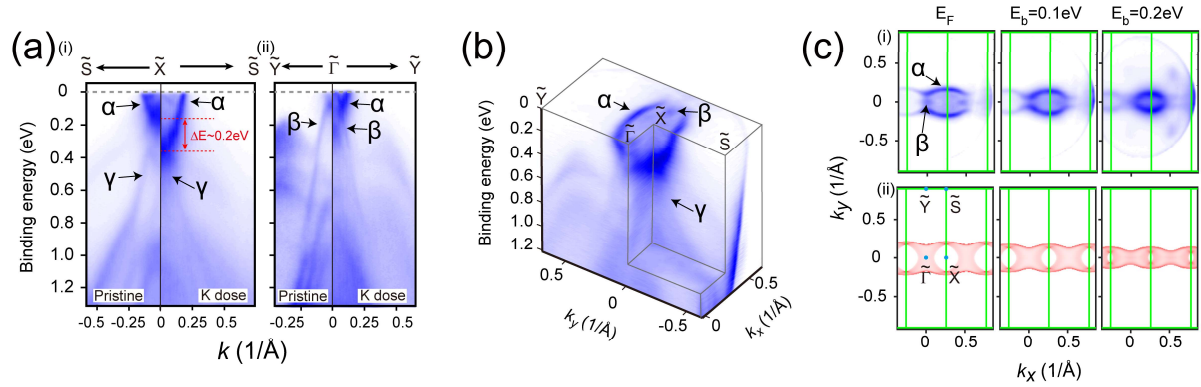

**FIG S9| Evolution of the electronic structure of TaSe$_3$ after potassium dosage.** (a) The comparison of the photoemission intensity of the spectrum measured along the (i) $\tilde{S} - \tilde{X} - \tilde{S}$ and (ii) $\tilde{Y} - \tilde{\Gamma} - \tilde{Y}$ directions. Left panels: dispersion of the pristine sample. Right panels: dispersions of the potassium dosed sample. Different bands are labeled for discussion. (b) 3D intensity plot of the photoemission spectra centered around $\tilde{X}$ of the potassium dosed sample. (c) (i) Photoemission spectral intensity map showing the constant energy contours at different binding energies. (ii) Corresponding calculated constant energy contours at the corresponding binding energies. The binding energy with a 200 meV shift was used to extract the correct constant energy contours from the calculation of the pristine TaSe$_3$ sample.

**Supplemental References:**